\documentclass[aps,prl,twocolumn,twoside,floatfix]{revtex4-1}
\usepackage{graphicx}
\usepackage{dcolumn}
\usepackage{bm}

\newcommand{\vect}[1]{\boldsymbol{#1}}

\newcommand{\ket}[1]{|#1\rangle}
\newcommand{\braket}[2]{\langle #1|#2\rangle}

\newcommand{\al}[0]{\rm{\boldsymbol{Al^+}}}
\newcommand{\mg}[0]{\rm{\boldsymbol{Mg^+}}}
\newcommand{\s}[0]{{}^1S_0}
\newcommand{\p}[0]{{}^3P_1}
\newcommand{\po}[0]{{}^3P_0}
\newcommand{\ii}[0]{(\rm{i})}

\newcommand{\dm}[0]{|\!\!\downarrow\rangle_{\rm{Mg}}}
\newcommand{\um}[0]{|\!\!\uparrow\rangle_{\rm{Mg}}}

\newcommand{\DB}[0]{\Delta_B}
\newcommand{\OB}[0]{\Omega_B}
\newcommand{\td}[0]{t_{d}}
\newcommand{\trsb}[0]{t_{\rm{rsb}}}

\newcommand{\wB}[0]{\omega_B}
\newcommand{\wOB}[0]{\omega_{0B}}
\newcommand{\wmg}[0]{\omega_{\rm{Mg}}}

\newcommand{\sss}[1]{\rm{\scriptscriptstyle{#1}}}

\begin{document}
\title{Trapped-Ion State Detection through Coherent Motion}

\author{D. B. Hume}
\altaffiliation{Present address: Kirchhoff-Institut f\"{u}r Physik, Universit\"{a}t Heidelberg, Im Neuenheimer Feld 227, 69120 Heidelberg, Germany}
\email[]{dhume@kip.uni-heidelberg.de}
\author{C. W. Chou}
\author{D. R. Leibrandt}
\author{M. J. Thorpe}
\author{D. J. Wineland}
\author{T. Rosenband}
\affiliation{Time and Frequency Division, National Institute of Standards and Technology, Boulder, Colorado 80305}
\date{\today}

\begin{abstract}
We demonstrate a general method for state detection of trapped ions that can be applied to a large class of atomic and molecular species.  We couple a ``spectroscopy'' ion ($\rm{^{27}Al^+}$) to a ``control'' ion ($\rm{^{25}Mg^+}$) in the same trap and perform state detection through off-resonant laser excitation of the spectroscopy ion that induces coherent motion.  The motional amplitude, dependent on the spectroscopy ion state, is measured either by time-resolved photon counting, or by resolved sideband excitations on the control ion.  The first method provides a simplified way to distinguish ``clock'' states in $\rm{^{27}Al^+}$, which avoids ground state cooling and sideband transitions.  The second method reduces spontaneous emission and optical pumping on the spectroscopy ion, which we demonstrate by nondestructively distinguishing Zeeman sublevels in the $^1S_0$ ground state of $\rm{^{27}Al^+}$.
\end{abstract}

\maketitle

Experiments on individual quantum systems face the challenge of isolating the system from the environment while making it accessible to external measurement and control. One way this conflict appears is during state detection when small energy differences between quantum states are amplified into directly measurable signals.  According to quantum theory, projective measurement leaves the system in its observed eigenstate, sometimes called a quantum nondemolition or nondestructive measurement.  However, unwanted perturbations to the quantum state make this ideal difficult to achieve experimentally, and near-perfect projective measurements, characterized by occasional ``quantum jumps'' between discrete signal levels, have been realized in only a few physical systems \cite{Blatt1988, Basche1995, Peil1999, Gleyzes2007, Lupascu2007}.

In the case of a trapped ion, ``electron shelving'' has become a standard technique used for projective measurements \cite{Blatt1988}.  Here, a transition between two atomic eigenstates acts as a switch for resonant photon scattering observed with a photon counter.  This requires a fast cycling transition at a suitable wavelength for laser sources and photon detectors, such that optical pumping does not disrupt the state being measured.  To overcome this limitation, quantum logic spectroscopy (QLS) was developed to indirectly detect the state of one or more ``spectroscopy'' ions by coupling them to a ``control'' ion of a different species \cite{Schmidt2006, Hume2007}.  The QLS detection sequence begins by cooling the ions close to the ground state.  Subsequent motional sideband transitions implement a quantum gate between the spectroscopy ion and the control ion whose state is then detected.  One drawback of QLS is the requirement of ground state cooling, which adds significant experimental complication.  Moreover, QLS cannot be applied in general to complicated atomic systems because it relies on narrow optical transitions in the spectroscopy ion, which must be accessible by CW laser sources.  For most ion systems such a resonance is unavailable or, through spontaneous emission, leads to depumping away from the measurement basis.

Here, we demonstrate state detection that does not rely on photon scattering and sideband transitions on the spectroscopy ion, making it suitable for a larger class of atomic and molecular systems.  As with QLS, laser excitation on the spectroscopy ion induces state-dependent motion that is detected using a control ion.  However, to avoid photon scattering, we apply only off-resonant interactions with the spectroscopy ion through a Stark shift $S^{\ii}$, dependent on the spectroscopy ion state $\ket{i}$.  Spatial variation of the Stark shift (i.e. due to an intensity or polarization gradient) gives rise to a state-dependent optical dipole force, ${\vect{F}^{\ii}(\vect{r},t) = - \vect{\nabla} S^{\ii}}$.  When the dipole force is modulated at the frequency $\omega_M$ of a normal mode of motion \cite{Monroe1996,Meekhof1996,Biercuk2011}, the ions behave like a classical driven harmonic oscillator with resonant driving force,  ${\vect{F}^{\ii}(t) = \vect{F}^{\ii}_0\cos(\omega_M t + \phi_M)}$, where $\vect{F}^{\ii}_0$ is the amplitude of the Stark--shift gradient at the ion position, and $\phi_M$ is its phase.

Due to resonant enhancement, the ion motion induced by this force can be detected with high sensitivity.  Recent work has demonstrated force detection by time-resolved photon counting at the level of 5 yN (${\rm{yN} = 1\times10^{-24}\, \rm{N}}$) \cite{Biercuk2010, Knunz2010}, equivalent to a Stark shift gradient of approximately 7.5 kHz over 1 $\rm{\mu m}$.  Furthermore, motional sideband transitions \cite{Monroe1996, Schmidt2006, Hume2007} allow for sensitivity to motion at the fundamental limit of single quanta.  Such high sensitivity can be exploited to make projective measurements by utilizing far off-resonant interactions with the spectroscopy ion that strongly suppress state transitions from photon scattering.

\begin{figure}
\includegraphics[width=\linewidth]{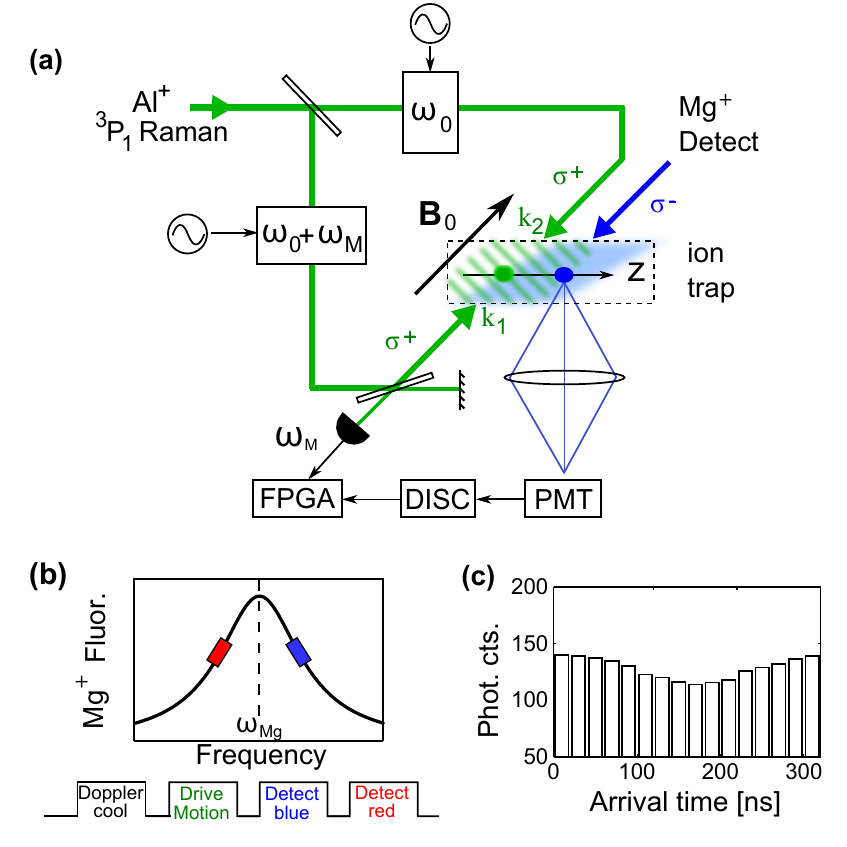}
\vspace{-18 pt}
\caption{(color online) (a) Experimental setup (DISC = discriminator, FPGA = field programmable gate array, PMT = photomultiplier tube).  Counter--propagating $\al$ $\p$ laser beams ($\lambda \simeq 267$~nm) excite ion motion at their difference frequency $\omega_M$, which is phase-stabilized with an interferometer near the ion trap.  With a laser tuned near resonance, the $\mg$ ion scatters photons that are counted and time-binned with respect to the driving force in an FPGA. All laser beams overlap both ions but because the $\al$ and $\mg$ wavelengths are substantially different, each beam interacts with only one species. (b) During detection, the $\mg$ beam ($\lambda \simeq 280$~nm) is tuned first above (blue, righthand rectangle) then below (red, lefthand rectangle) the optical resonance, which amplifies then damps the motion respectively. Fluorescence from $\mg$ is modulated by the ion motion with the modulation phase shifted by $\pi$ between the blue and red pulses. (c) Peak-to-peak $\mg$ fluorescence modulation of $20 \%$ when $\al$ is in the $\s$ state.  In the $\po$ state modulation is absent.  The motional period of $2 \pi/\omega_M = 340$~ns is divided into 16 photon-arrival-time bins.  \label{fig:ExpDiagram}}
\end{figure}

The experimental system, sketched in Fig.~\ref{fig:ExpDiagram}(a), confines a $^{25}\mg$ ion and a $^{27}\al$ ion along the axis ($\vect{\hat{z}}$) of a linear Paul trap whose normal mode frequencies for a single $\mg$ ion are $\{\omega_x, \omega_y, \omega_z\} = 2\pi\times\{5.1 , 6.8, 3.0\}$~MHz.  The ion pair Mg-Al has an equilibrium separation of 3.1~$\mu$m, with a quasi-center-of-mass (COM) \footnote{Here COM refers to an approximate center-of-mass mode, in which the ions oscillate in phase but with slightly different amplitudes due to their unequal masses} mode frequency of $\omega_{M}= 2\pi\times 2.94$~MHz, corresponding to a spread in the ground-state wavefunction of ${z_{\sss{0,Mg}} = 5.64}$~nm and ${z_{\sss{0,Al}} = 5.86}$~nm for $\mg$ and $\al$ respectively.  The optical dipole force is produced by two $\sigma^+$-polarized, counter--propagating laser beams, which interfere at their focus ($\rm{\simeq 40\, \mu m}$ diameter) to produce a ``walking'' wave intensity pattern \cite{Monroe1996}.  The beams are detuned from the $\s\rightarrow\p$ transition (linewidth $\Gamma/2\pi =$ 520 Hz) \cite{Trabert1999} in $\al$ by $\Delta_R/2\pi \simeq 20$~MHz and have a frequency difference $\Delta \omega = \omega_M$.  To maintain phase coherence between the lasers and the ion motion, we measure the phase difference of the beams in an interferometer near the trap and stabilize it with an acousto-optic modulator in one of the beam paths.  We measure a relative coherence time at the position of the ions of about 200 s.

Our first experiment implements the excitation/detection sequence depicted in Fig.~\ref{fig:ExpDiagram}(b). It begins by laser cooling all normal modes to near the Doppler limit.  Next, the optical dipole force of duration $t_d = 250\, \rm{\mu s}$, applied to the $\al$ ion, excites the COM mode along the trap axis.  We detect the induced harmonic motion by observing the $\mg$ ion's oscillating velocity and resulting Doppler-shift as modulation of the photon-scattering rate when the laser is tuned to a slope of the ion resonance [see Fig. 1b].  The $\sigma^-$-polarized $\mg$ detection beam is focussed to a 60 $\rm{\mu m}$ diameter, propagates in a direction anti-parallel to the magnetic field (at $45^\circ$ to the $z$-axis) and is nearly resonant with the ${|{3s\,^2S_{1/2}}, F = 3, m_F=-3\rangle}$ ${\rightarrow |{\rm 3p\,^2P_{3/2}}, F = 4, m_F=-4\rangle}$ cycling transition (frequency $\wmg$, linewidth ${\Gamma_{\sss{Mg}} = 2\pi\times41.4}$~MHz).  We collect fluorescence photons with an efficiency of 0.4~\% during two laser pulses of intensity 3 kW/m$^2$.  The first blue-tuned pulse ($\omega_{\rm{blue}} \simeq \wmg + \Gamma_{\sss{Mg}}/2$) amplifies the ion motion while the second red-tuned pulse ($\omega_{\rm{red}} \simeq \wmg - \Gamma_{\sss{Mg}}/2$) damps it \cite{Vahala2009}.  Fluorescence during both pulses is modulated at the frequency $\omega_M$, albeit with a $\pi$ phase shift that is compensated electronically.  We choose the respective pulse durations ($t_{\rm{blue}} = 400\, \rm{\mu s}$, $t_{\rm{red}} = 200\, \rm{\mu s}$) empirically to maximize the detection signal.

\begin{figure}
\includegraphics[width=\linewidth]{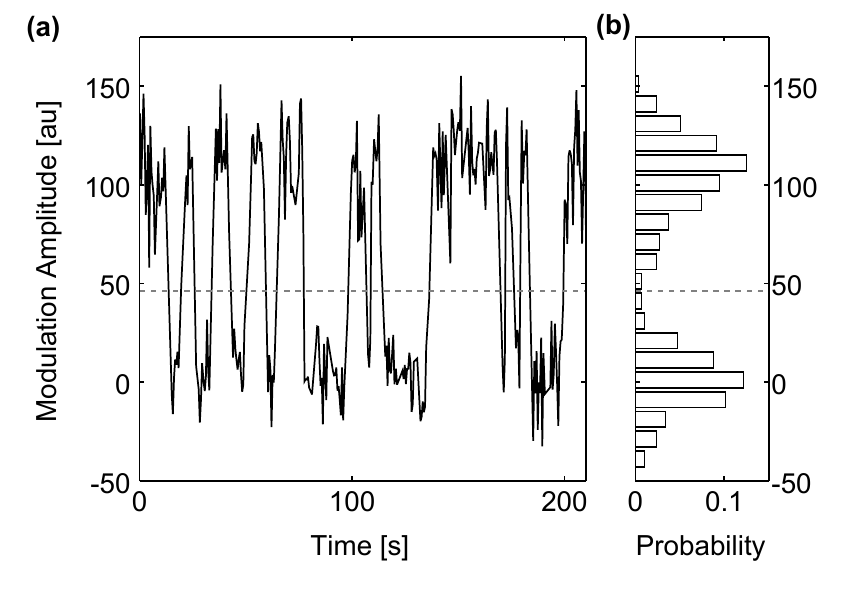}
\vspace{-24pt}
\caption{(a) Steps in fluorescence modulation amplitude corresponding to quantum jumps between the $\s$ and $\po$ clock states in $\al$, which are induced by a near-resonant laser beam or by spontaneous decay.  The amplitude is extracted at the calibrated modulation phase and averaged over 0.53 s.  (b) Histogram of mean modulation amplitudes from the entire data set.  Dashed lines indicate the threshold for detecting transitions. \label{fig:ClockDetect}}
\end{figure}

The above detection sequence is applied repetitively to determine the $\al$ state.  A sinusoidal fit to the time-binned $\mg$ photon counts yields the modulation amplitude at the calibrated frequency and phase of ion motion [Fig.~\ref{fig:ExpDiagram}(c)].  We observe 20 \% peak-to-peak modulation relative to the mean fluorescence when $\al$ is in the $\s$ state.  Figure~\ref{fig:ClockDetect}(a) shows the modulation amplitude as a function of time with periodic laser pulses inserted to probabilistically drive $\s\leftrightarrow\po$ ($\lambda_{^3P_0} = 267.0$ nm).  The presence of modulation in the ion fluorescence corresponds to the $\al$ ion occupying the $\s$ ground state, while its absence corresponds to the ion in the $\po$ state, which has negligible interaction with the $\s\leftrightarrow\p$ laser beams ($\lambda_{^3P_1} = 267.4$ nm).   The quantum jumps visible in the data are caused either by the periodic laser pulses, or by spontaneous decay from the $\po$ state.   Figure~\ref{fig:ClockDetect}(b) shows a histogram of the modulation amplitude; these data reach 93 \% state-detection fidelity within 80 ms, which could be improved by higher modulation amplitudes or photon collection efficiency.

\begin{figure*}[ht!]
\includegraphics[width=\textwidth]{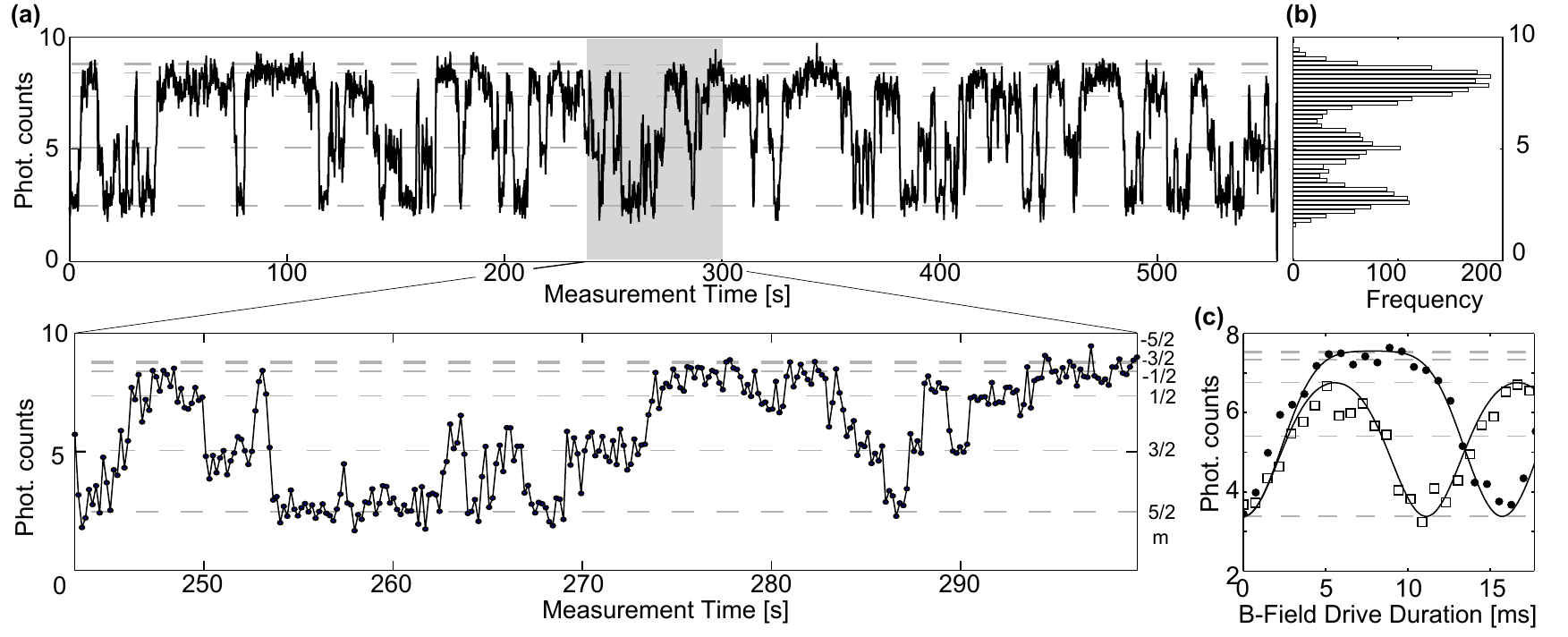}
\vspace{-20 pt}
\caption{(a) Quantum jumps between Zeeman substates of the $\s$, $I=5/2$ manifold in $\al$. Points record the fluorescence levels in bins of 120 consecutive detections (1.6 ms detection duration).  Dashed lines show predicted fluorescence levels based on the calibration for $m = +5/2$. \emph{top} Several minutes of repeated detection cycles. \emph{bottom} A detailed view of 50 seconds of the detection data. (b) A histogram of the ion fluorescence signal (mean taken in bins of 120 experiments). (c) Magnetic resonance experiment an $^{27}\al$ ion initialized in the $\s$, $m = 5/2$ state with $\DB=0$ (filled circles) and $\DB = 2\OB$ (open squares).  The solid curves are calculations of fluorescence based on a fitted value of $\Omega_B$, and an initial thermal motional mode with $\langle n\rangle = 0.15$.   Dashed lines show the expected detection fluorescence levels for different Zeeman states. \label{fig:ZeemanJump}}
\end{figure*}


In a second experiment, similar to a proposal for molecular-state detection~\cite{Koelemeij2007}, we explore a more sensitive method for distinguishing small amplitudes of coherent motion using resolved sideband transitions on the control ion.  In this method, two hyperfine states of $^{25}\mg$ separated by 1.8 GHz form a quantum bit ${\dm\equiv|^2S_{\frac{1}{2}}, F=3, m_F=-3\rangle}$ and ${\um\equiv|^2S_{\frac{1}{2}}, F=2, m_F=-2\rangle}$ as described in previous work \cite{Hume2009}.  During a detection cycle, the ions are first cooled by stimulated-Raman transitions to near the ground state of motion for the axial modes \cite{Monroe1995} before applying the same optical dipole force described above.  This produces a coherent state of motion $\ket{\beta^{\ii}}$ with probability distribution $p^{\ii}(n) = e^{-|\beta^{\ii}|^2}|\beta^{\ii}|^{2n}/(n!)$ in terms of Fock states, $\ket{n}$.  We probe this distribution with a red--sideband (RSB) pulse on the $\mg$ qubit transition \cite{Meekhof1996}, during which the $\dm$ state probability evolves as $P^{\ii}_{\downarrow}(\trsb) = \sum_{n}p^{\ii}(n)\cos^2(\Omega_n \trsb)$.  Here, $\Omega_n$ is the RSB Rabi flopping rate for ${\dm\ket{n}\rightarrow\um\ket{n-1}}$ and we have ignored relaxation of the $\mg$ qubit which is a small effect for typical pulse durations.  A final $\mg$ resonance fluorescence detection pulse distinguishes $\um$ and $\dm$.  By repeating this detection cycle several times we measure $P_{\downarrow}$ for a particular RSB pulse duration $\trsb$ and gain information about $\beta$ to identify the $\al$ state.

To demonstrate this method, we experimentally distinguish Zeeman substates of the $\al$ $\s$ ground state.  The applied static magnetic field $\vect{B}_0$ gives rise to six Zeeman levels $|I,m\rangle$ for $m=-5/2$ to $+ 5/2$ that, in the absence of other perturbations, are equally separated in energy by $\hbar \wOB = |g_I \mu_B B_0|$, where $g_I=-0.00097248(14)$ is the nuclear $g$-factor \cite{Rosenband2007} and $\mu_B$ is the Bohr magneton.  During the detection sequence, the optical dipole force leaves the ions in motional state $\beta^{(m)}$ dependent on the state $|I,m\rangle$. The primary factor affecting $\beta^{(m)}$ is the Clebsch-Gordan coefficient $C_m\equiv \braket{\frac52, m;1, 1}{\frac72,m+1}$ that describes the relative strength of the electric dipole coupling between $|I,m\rangle$ and $\ket{\p, F=7/2, m_F = m+1}$.  Also, the overall detuning, ${\Delta_{m} \equiv \Delta_R + g_{\p}(5/2 - m)\mu_B B_0}$ is shifted by the linear Zeeman effect on the $\p$ state.  Here, $g_{\p} \simeq 3/7$ is the $\p$ $g-$factor, when nuclear and relativistic corrections are neglected.  For small $\al$ motional amplitudes in the Lamb-Dicke limit ${(\beta z_{\sss{0, Al}} \ll \lambda_{^3P_1} / 2 \pi \sqrt{2})}$,
\begin{equation}
 |\beta^{(m)}| = \frac{\eta \td \left(\Omega'_0 C_m\right)^2}{|\Delta_m|},
 \label{eq:Beta}
\end{equation}
where $\Omega'_0$ is the carrier Rabi rate for the ${|I,m = 5/2\rangle \rightarrow |{\p}, F = 7/2, m_F = 7/2\rangle}$ transition, and $\eta = 2 \pi \sqrt{2}z_{\sss{0, Al}}/\lambda_{^3P_1}$ is the Lamb-Dicke parameter.

We calibrate $\Omega'_0$ by optically pumping to ${|I,m = 5/2\rangle}$, applying the optical dipole force for duration $\td$, then probing the Fock state distribution with a $\mg$ RSB pulse of variable duration \cite{Meekhof1996}.   We fit a coherent state amplitude $\beta^{(5/2)}$ to the resulting curve and calculate the other $\beta^{(m)}$ based on Eq.~\ref{eq:Beta}.  In the experiment here, with $\Omega'_0/2\pi = 0.85$~MHz and $t_d = 50$~$\rm{\mu s}$, $\{\beta^{(m)}\} = \{0.05, 0.16, 0.37, 0.71, 1.26, 2.15\}$ in order of increasing $m$.  We choose $\trsb = 2.8$~$\rm{\mu s}$ to differentiate several of the Zeeman states for a typical Rabi-flopping rate $\Omega_{n=1}/2\pi = 0.07$~MHz.

In Fig.~\ref{fig:ZeemanJump}(a) we record quantum jumps of the fluorescence signal as a function of time when we repeatedly apply the detection procedure without state preparation.  Each data point corresponds to the average photon counts from 120 consecutive detection cycles (1.6 ms cycle time).  Several distinct fluorescence levels are visible, which agree with the above calibration.  Jumps in the fluorescence correspond to changes of the Zeeman state that occur during the measurement process.  The jumps appear to be caused by polarization imperfection in the $\p$ Raman beams.  Here, residual $\pi$ or $\sigma^-$-polarization can induce a two-photon stimulated-Raman process between Zeeman states.  We estimate that a laser field with 3 \% residual $\pi$-polarization would cause the observed rate of quantum jumps.

The tendency of the fluorescence to persist at a particular level indicates the nondestructive nature of the detection method.  This is possible without a cycling transition for the Zeeman sublevels because the off-resonant laser pulses give only a small probability ($\approx 1\times10^{-4}$ for $m = 3/2$) of spontaneous-Raman scattering through the $\s\leftrightarrow\p$ transition in a single detection cycle.  In this regime, a histogram of averaged fluorescence levels (Fig.~\ref{fig:ZeemanJump}(b)) for the entire detection record exhibits separate maxima in the distribution, which correspond to the resolvable states.

To further characterize this detection protocol, we perform a nuclear magnetic resonance experiment on the $\s$ electronic ground state of $^{27}$Al$^+$.  An oscillating magnetic field $B_1 \cos{\wB t}$ in a direction perpendicular to the static field $\vect{B}_0$ induces transitions between Zeeman states.   Our case of small detuning ($\DB \equiv \wB - \wOB \ll \wB$) and weak drive ($B_1<B_0$), allows the rotating--wave approximation, where the angular momentum is affected primarily by the component of the oscillating field that rotates about $\vect{B}_0$ in the same sense as the Larmor precession of the magnetic dipole.  In the general case we solve the coupled Schr\"{o}dinger equations to determine the probability $P_{I,m}(t)$ of finding the ion in state $|I,m \rangle$.

We initialize $\al$ in $|I,m = 5/2\rangle$ and plot the fluorescence signal for different values of $\DB$ as a function of the $B$-field modulation pulse duration (Fig.~\ref{fig:ZeemanJump}(c)).  In the experiment $B_0 = 0.74$~mT ($\wOB \simeq 2\pi\times8.3$~kHz).  A fit to the data for zero detuning $(\Delta_B=0)$ yields the Rabi flopping rate $\Omega_B$, and this value is used to calculate the curve for $\DB = 2\OB$ with no additional free parameters.  Uncertainty associated with depumping during detection and imperfect ground state cooling affects the agreement between theory and experiment.  For the theoretical curves shown here we use a residual (thermal) Fock state occupation of $\langle n\rangle = 0.15$ based on a separate calibration from sideband measurements after cooling.

In summary, we have explored a general method for detection of quantum states of trapped ions and have experimentally implemented two specific protocols to detect states in $\al$ by exciting state-dependent coherent motion.  Although demonstrated on single ions, the same approach can be used to detect the states of multiple ions, and the optical driving force can be created in a variety of configurations with respect to laser frequencies, geometry, and polarization.  As one practical application, the modulated fluorescence method can simplify $\al$ optical clocks \cite{Rosenband2007} because lasers for ground-state cooling and resolved--sideband transitions are no longer needed.  Technical improvements such as higher photon-collection efficiency and further optimized laser pulses can increase state-detection efficiency.  Electromagnetically-induced transparency \cite{Roos2000} may also improve efficiency, because the reduced atomic line-width in this configuration enhances the velocity-sensitivity of the ion fluorescence.

Furthermore, the resolved--sideband method provides greater sensitivity to small amplitudes of motion that are generated by weak, non-destructive, interaction between the Stark-shifting lasers and the spectroscopy ion.  Because internal state changes from spontaneous-Raman scattering can be suppressed, the present technique approaches the textbook ideal of a quantum measurement, where state collapse is the only effect.  This provides a new route to perform spectroscopy on ion species where state changes from optical pumping are problematic.  For example, the method might be used to detect ro-vibrational resonances in molecular ions \cite{Vogelius2006, Koelemeij2007} by observing the magnitude and phase of coherent motion excited by far-off-resonant lasers.

\begin{acknowledgments}
We thank C.~Oates for contributions to this work, and acknowledge helpful comments from U.~Warring and R.~J\"ordens.  The work was supported by AFOSR, the DARPA QuASaR program, ONR and IARPA.
\end{acknowledgments}


%

\end{document}